\documentclass[aip,jcp,amsmath,amssymb,reprint,nofootinbib]{revtex4-1}

\usepackage{graphicx}
\usepackage{gensymb}
\usepackage{eqnarray,amsmath}
\usepackage[T1]{fontenc}
\usepackage{scrextend}

\usepackage{color}

\begin{document}

\title{Sequential Snapping and Pathways in a Mechanical Metamaterial}%

\author{Jiangnan Ding}
\email{jiangnanding@gmail.com}
\author{Martin van Hecke}
\email{mvhecke@gmail.com}
\affiliation{Huygens-Kamerlingh Onnes Lab, Leiden University, PObox 9504, 2300 RA Leiden, The Netherlands \\
	and  AMOLF, Science Park 104, 1098 XG Amsterdam, The Netherlands}

\begin{abstract}
Materials which feature bistable elements, hysterons, exhibit memory effects. Often these
hysterons are difficult to observe or control directly. Here we introduce a mechanical metamaterial in which slender elements, interacting with pushers, act as mechanical hysterons. We show how we can tune the hysteron properties and pathways under cyclic compression by the geometric design of these elements and how we can tune the pathways of a given sample by tilting one of the boundaries. Furthermore, we investigate the effect of the coupling of a global shear mode to the hysterons,
as an example of the interactions between hysteron and non-hysteron degrees of freedom. We hope our work will inspire further studies on designer matter with targeted pathways.
\end{abstract}

\date{\today}%
\maketitle

\section{Introduction}

Hysteretic elements commonly occur in complex materials and play a key role in the understanding of memory effects \cite{mungan2019structure,mungan2019networks,preisach1935magnetische,regev2021topology,jules2021delicate,lindeman2021multiple,keim2021multiperiodic,martinPRE,keim2019memory,yoavpre,hadrien,terzi2020state,forte2021muaori,yang2019multi,rafsanjani2015snapping,sun2019snap,yang20201d}.
Intuitively, when cyclically driving a complex system, one imagines these elements to undergo sequences of flipping transitions associated with hopping between metastable states. To understand these sequences, it is often possible to model
these elements as hysterons: hysteretic elements which
which flip their internal state $s$ from '0' to '1' when the local driving exceeds the upper switching field $\varepsilon^+$, and which flip from '1' to '0' when the driving falls below the lower switching field $\varepsilon^-$ (Fig.~1a) \cite{preisach1935magnetische,mungan2019networks,
lindeman2021multiple,keim2021multiperiodic,martinPRE,keim2019memory}.
By specifying the values of the switching fields of a collection of hysterons,
and potentially their interactions, one can determine the transitions between
all collective states $S:=\{s_1,s_2,\dots\}$, and represent these
in a transition graph (t-graph) which takes the form of a directed (multi)graph \cite{martinPRE,mungan2019structure,lindeman2021multiple,keim2021multiperiodic,mungan2019networks,hadrien,paulsen2019minimal}.
The (topological) organization of such t-graphs
characterize the complex response of complex media and in particular
memory effects such as Return Point Memory (RPM), transient memories, and subharmonic response \cite{mungan2019networks,regev2021topology,jules2021delicate,paulsen2019minimal,mungan2019structure,lindeman2021multiple,keim2021multiperiodic,martinPRE,terzi2020state,deutsch2004return,goicoechea1994hysteresis}.

Controlling, characterizing and manipulating such hysterons is challenging in disordered systems such as crumpled sheets and amorphous media \cite{paulsen2014multiple,keim2020global,adhikari2018memory,matan2002crumpling,lahini2017nonmonotonic,yoavpre,hadrien}. Here we propose instead to leverage the design freedom of mechanical metamaterials to embed hysterons into a flexible metamaterial \cite{overvelde2015amplifying,bertoldi2017flexible,forte2021muaori,yang2019multi,rafsanjani2015snapping,sun2019snap,yang20201d}.
 This allows to control and tune their switching fields and to directly observe the sequences of hysteron flippings that constitute the deformation pathways and yield the t-graph. Developing such metamaterial platforms is an important step towards achieving
materials with deformation pathways on demand \cite{Coulais,mungan2018}, with targeted memory properties and specific responses to cyclical driving, and with advanced pathways that include elementary computations. Moreover, such metamaterials allow to explore
generality and robustness of the hysteron picture, and to more closely explore the material properties (e.g. sensitivity to boundary conditions) of multistable materials. Finally, metamaterials allow
to explore the interactions between non-discrete degrees of freedom, for example given by visco-plastic relaxation and solid-on-solid friction, which may lead to additional timescales and continuous degrees of freedom not considered in simple hysteron models \cite{yoavpre}.

Here we introduce a simple metamaterial platform in which mechanical hysterons with controllable switching fields can be embedded. We start from the well-known
biholar metamaterials, which translate global uniaxial compression to local rotation and compression \cite{florijn1,florijn2,zhang2019ordered}. We then leverage the hysteretic snapping of beams between left- and right buckled states to locally replace slender elements of the metamaterial by hybrid pusher-beam elements\cite{bertoldi2017flexible,forte2021muaori,yang2019multi,rafsanjani2015snapping,sun2019snap,yang20201d,zhang2020tunable}.
We show that tuning their design parameters allows to access qualitatively different pathways. Moreover, we use gradients in the boundary conditions to independently tune the effective switching fields of the hysterons, thus obtaining multiple pathways from a single sample \cite{hadrien}.
Finally, we show how subtle frictional effects allow to slowly evolve the switching fields, giving rise to a history dependent response beyond that captured by simple hysteron models. Together, our work opens new directions for the experimental study and control of multistable materials.

\section{(Un)snapping in biholey metamaterial}\label{sec:OneBeam}

\begin{figure}[ht]
	\includegraphics[width=.9\linewidth]{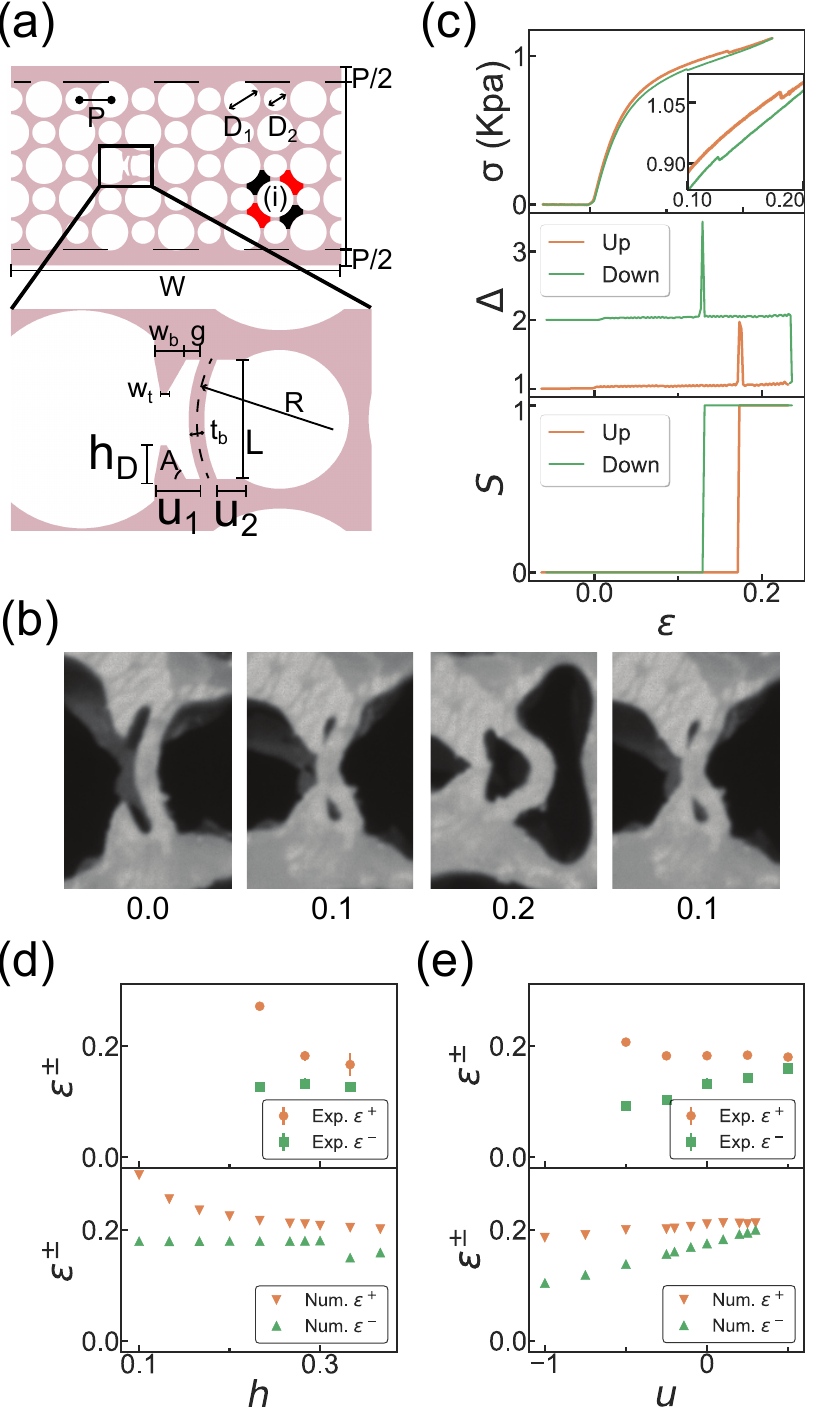}

\caption{
		(a) Quasi-2D biholey metamaterial of height $H=50$mm, width $W=100$mm and thickness $T=18$mm. The hole pattern has pitch $P=10$mm, the holes have diameters $D_1=11$mm and $D_2=7$mm.
 The unit cell (i) of the biholey metamaterial  consists of four  diamond shaped islands  that highlighted in black and red and hinges connecting islands.
Zoom-in: geometry of the defect beam (not to scale). Here,
$L=6$mm, and we fix the dimensionless radius of curvature $r:=R/L=0.8$, dimensionless thickness $t:=t_b/L=0.13$, and dimensionless location $u:=\frac{u_1 - u_2}{u_1 + u_2}=0$.
The pushers are characterized by $\{g,A,w_t,w_b,h:=h_D/L\}$ = \{0.7 mm, 60$\degree$, 0.3 mm, 1.6 mm, 0.28 \}.
(b) Example of the evolution of the defect beam geometry under cyclic compression, showing the pushers getting  into contact of the beam and initiating a snapping towards the right.
(c) Stress $\sigma$, image differences $\Delta$ and hysteron state $s$ as
function of strain $\varepsilon$ for the sample in (a-b). Note that the image differences for
compression and decompression are offset by one for visibility.
(d-e) Critical switching strains $\varepsilon^{\pm}$ as function of the dimensionless pusher height $h$ and beam position $u$. We have performed three independent runs on two samples, and calculated our
errorbar based on these six datasets.}
	\label{fig1}
\end{figure}

We use quasi-2D biholey metamaterials, in which we embed one or more structures which act as mechanical hysterons. The biholey design consists of alternating smaller and larger holes, separated by alternating precurved beams which are connected in groups of four in diamond shaped islands (Fig.~1a) \cite{florijn1,florijn2,zhang2019ordered,bertoldi2010negative,mullin2007pattern,bertoldi2008mechanics,johnson2017buckling,overvelde2014relating,yang2016phase,zhang2019programmable}. Under vertical compression, the vertical beams curve even more, and the diamond shaped islands exhibit counter-rotating motion \cite{florijn1,florijn2,zhang2019ordered,bertoldi2010negative,mullin2007pattern,bertoldi2008mechanics,johnson2017buckling,overvelde2014relating,yang2016phase,zhang2019programmable}. We focus here on biholey metamaterials of consisting of 9 $\times$ 5 holes (Fig.~1a).

Each hysteron is composed of a 'defect' beam with curvature opposite to that expected in the biholar design, and two adjacent pushers (Fig.~1b). This design ensures 	that under compression the defect beam will be pushed 	by these pushers from a left curved to a snapped, right curved state.
Under decompression, the defect beam then 	snaps back to its left curved state.
 As the defect beams are strongly pre-curved, they do not exhibit buckling.
We label the unsnapped and snapped states of the defect beam as '$s=0$' and '$s=1$', to stress that the defect beam acts as a mechanical hysteron \cite{jules2021delicate,hadrien}.

The properties of the mechanical hysterons depend on the design of both the defect beam and the triangular pushers, as well as on its
location in the metamaterial (Fig.~1a-b). The defect beam is specified by the
length $L$, its dimensionless radius of curvature $r:=R/L$, dimensionless thickness $t:=t_b/L$, and dimensionless horizonal location $u:=\frac{u_1 - u_2}{u_1 + u_2}=0$, where the defect beam is closer to the smaller (larger) holes when $u > 0$ ($u$ < 0). The pushers are characterized by the angle $A$, the width of their bases $w_t$ and $w_b$, their dimensionless height $h:=h_D/L$, and the gap $g$ between pusher and beam (Fig.~1a-b).
Based on exploratory experiments and finite element simulations,
we fix  the beam parameters $\{L, r, t\} =$ \{6 mm,  0.8, 0.13 \} , fix the pusher parameters $\{g, A, w_t, w_b\} =$ \{0.7  mm, 60$\degree$, 0.3 mm, 1.6 mm  \}, and vary the dimensionless pusher height $h$ and horizontal beam location $u$.

We now first explore the behavior of a single mechanical hysteron
(Fig.~\ref{fig1}(b-c)).
We apply cyclic loading using a (de)compression rate of  0.2 mm/sec,
which leads to nearly quasistatic behavior --- much faster rates lead to inertial effects, while much lower rates lead to creep effects.
Under compression by a strain
$ E_y / H$, where $E_y$ denotes the displacement of the compression plate,
the defect beam shown in Fig.~1c will initially ($\varepsilon \lesssim 0.18$)
bend left. In contrast,
the rotation of the diamond shaped islands makes the
tips of the pushers move right, so that they eventually come into contact with the defect beam. Further compression then causes a hysteretic transition of the defect beam into a right-snapped state at $\varepsilon^{+} =  0.18$. To detect this transition we use difference imaging, which is sensitive to sudden motions, and define $\Delta$ as normalized difference of the snapshots of the defect snappers (see Supplementary Material).
We use a window around each defect beam of 	10$\times$20 mm to determine $\Delta$.
We note that the snapping behavior, visible as a small but sharp drop in the compressive stress $\sigma:=
F/(W T)$, where $F$ is the compressive  force, can be seen very clearly in the image differences, and indicates that the hysteron state, $s$, switches from 0 to 1, as shown by  the orange curve in Fig.~\ref{fig1}(c).
Under decompression, the defect beam then snaps back to its left curved state
at $\varepsilon^{-} = 0.13$, which can be seen in both the stress signal and the image differences  (note that we have offset the image differences of the downsweep for clarity).
We note that the $\varepsilon^{+}$ is larger than $\varepsilon^{-}$, so that there is a bistable region, as expected for a hysteretic transition (Fig.~1c).

We can modify the characteristic (un)snapping strains, $\varepsilon^{\pm}$, by modifying the
design of the snappers, and focus on the role of the height of the pusher
$h$ and beam position $u$. We observe that $\varepsilon^+$ increases
for lower heights $h$ --- more compression is needed to induce a snapping event with smaller pushers --- whereas $\varepsilon^-$ is essentially independent of $h$ --- which makes sense, as before unsnapping, the defect beam and pusher are not in contact (Fig.~\ref{fig1}(d)).
We note that our numerical results show a discontinuity  for $h\sim 0.32$ --- here the pushers are so high that two opposing pusher come into contact, hinder rotation of the diamonds above and below the defect beam, and delay the unsnapping transition.
As a function of the relative defect beam position $u$, we observe that $\varepsilon^+$ is nearly constant, while $\varepsilon^-$ increases with $u$ (Fig.~\ref{fig1}(e)). We interpret this trend as follows: due to rotation of the diamonds, the effective distance between top and bottom of the defect beam decreases when $u$ is decreased; such defect beams are thus more compressed, and unsnap for lower values of $\varepsilon$. Finally we note that
for extreme parameter choices, instead of snapping, the beam undergoes smooth deformations
and stops acting as a hysteron; this occurs for example when $u>0.5$. We conclude that the geometric parameters of the snapping beam allow to tune the upper and lower switching fields $\varepsilon^\pm$ of the corresponding hysteron.

\section{Transition pathways and states}\label{sec:pathway-notilt-sampleA}
\begin{figure}[ht]
	\includegraphics[width=0.8\linewidth]{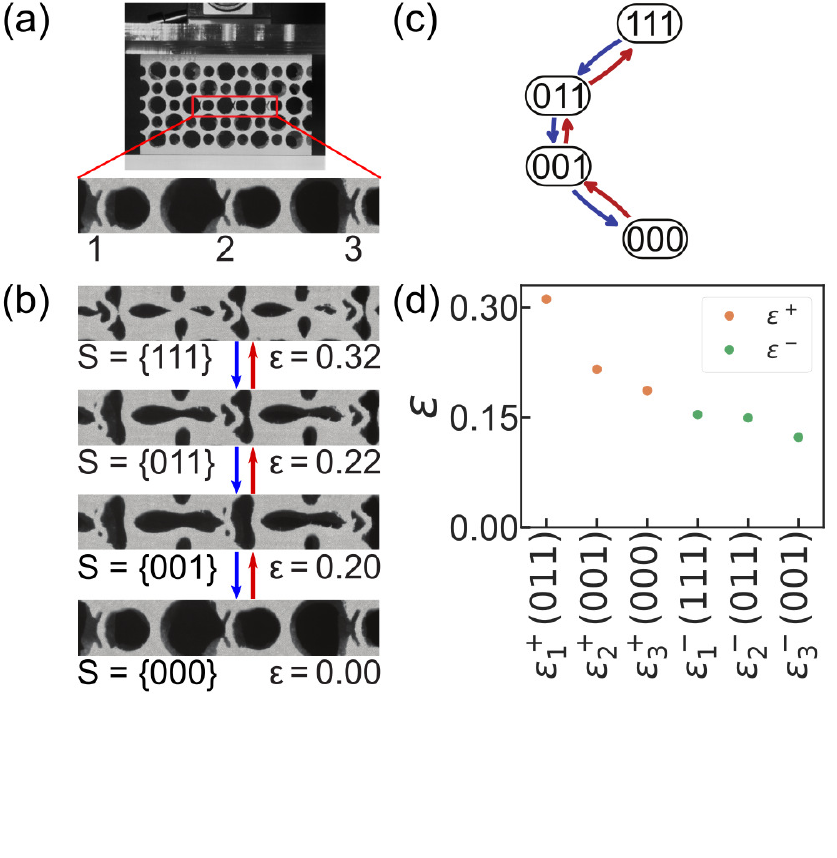}
	\caption{		Robust pathways in sample A.
		(a) Sketch of the sample with three defects labeled $1$, $2$ and $3$ respectively.
		(b) Snapshots of the defect beams during a compression/decompression cycle, showing the distinct states.
		(c) The transition graph of the sample A.
		(d) The switching fields $\varepsilon^\pm$ A ordered from large to small.			 }
	\label{fig2}
\end{figure}

We now explore the transition pathways in a metamaterial with three defect beams under cyclic compression (Fig.~\ref{fig2}). We label the
defects beams as $1$, $2$, and $3$ (left to right), their individual states as $s_1$, $s_2$ and $s_3$, and their collective state as $S:=\{s_1,s_2,s_3\}$.
In the absence of interaction, the pathways are determined by the relative ordering of the upper and lower switching strains of each hysteron \cite{mungan2019structure,terzi2020state}, which we denote by $\varepsilon_i^\pm$, where the  subscript $i$  labels the switching hysteron; if there are interactions, the situation can become more complex, and we denote the switching fields as
$\varepsilon^\pm_i(S)$, where $S$ is the state just before the transition \cite{martinPRE,hadrien,lindeman2021multiple,keim2021multiperiodic}.

We first aim to design a metamaterial with the simplest possible pathway,
such that under compression we observe a pathway $\{000\}\!\rightarrow\!\{001\} \!\rightarrow\!\{011\}\!\rightarrow\!\{111\}$, and we visit the same states in opposite order under decompression. Assuming that interactions can be ignored, this requires $\varepsilon_1^+ > \varepsilon_2^+ >  \varepsilon_3^+$ and
$\varepsilon_1^- > \varepsilon_2^- >  \varepsilon_3^-$  \cite{martinPRE,lindeman2021multiple,keim2021multiperiodic}.
We thus chose design parameters for our hysterons consistent with this ordering  ---  the upper switching fields are mostly controlled by $h$ and decrease for increasing $h$ (Fig.~1(d)), and we choose dimensionless heights
$\{h_{1}, h_{2}, h_{3}\} =  \{ 0.23, 0.28, 0.33\}$; the lower switching fields are mostly controlled by $u$ and increase with $u$, and we chose
$\{u_{1}, u_{2}, u_{3}\} =  \{0.2, 0, -0.1\}$.
We refer to this as sample 'A'.

Performing
cyclic compression and decompression, we observe the targeted pathway in sample A (Fig.~2(b)). We can collect the states and their transitions in a very simple transition graph (t-graph), where we denote the different states as nodes, and the 'up' transitions under compression, and 'down' transitions under decompression by red and blue arrows (Fig.~2(c)) \cite{mungan2019networks,paulsen2019minimal,mungan2019structure,martinPRE,terzi2020state}.
 Finally, each transition is associated with a specific value of the compression strain. We denote these as $\varepsilon^\pm_i(S)$, where the superscript $\pm$ denotes up or down transitions, the subscript $i$ the label of the switching hysteron, and $S$ the initial state, just before the transition, and show these in Fig.~2(d). We note that although $\varepsilon^-_3 (111) \approx 0.154$ and $\varepsilon^-_2 (110) \approx 0.149$ are quite close,
the ordering of the switching fields is consistent with our targeted ordering, and our observed pathway is robust.

\section{Tuning pathways by tilting}\label{tilt}
\begin{figure}[t]
	\includegraphics[width=\linewidth]{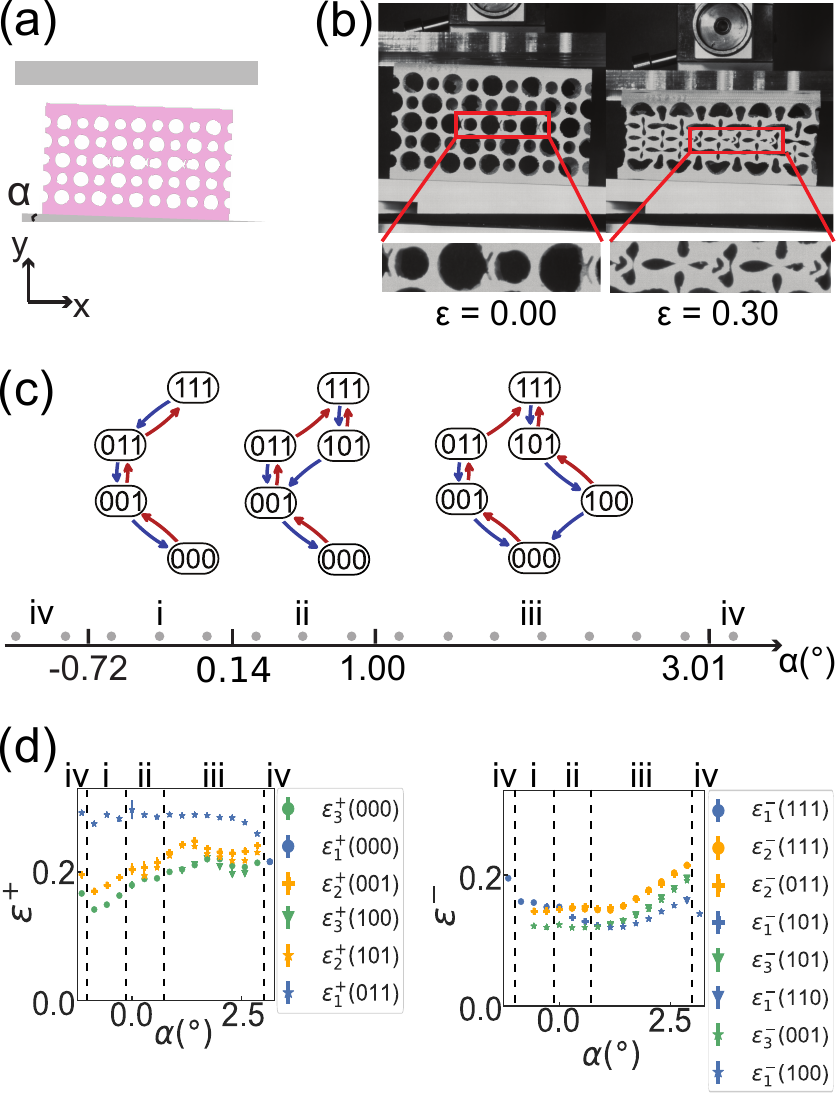}
	\caption{Tuning pathways by tilting.
		(a) Sketch of a sample where the bottom boundary is tilted over an angle $\alpha$.
		(b) Snapshots of the sample A tilted by $\alpha=1.7\degree$ at strains as indicated; notice the emergence of shear in the inset of the right panel.
		(c) As function of $\alpha$, we observe three distinct pathways;
gray dots indicate tilt angles where we determined the pathways, and the boundaries
are estimated by bisection.
		(d) Corresponding critical switching fields as function of $\alpha$ and state $S$. As the switching fields vary weakly with state, the corresponding symbols nearly overlap. The error bar is of the order of the symbol size and denotes the spread of $\varepsilon^{\pm}_i$ of repeat compression.} 	
	\label{fig3}
\end{figure}

Spatial gradients in the driving can modify
the relation between local and global driving magnitude, and allow a relative shift of the switching fields of adjacent hysterons \cite{hadrien}. Here we
use tilting of the bottom boundary in our experiments
to modify the transition pathways of a given sample (Fig.~3a-b). To do so, we employ a bottom plate with an adjustable tilt angle $\alpha$ (Fig.~\ref{fig3}).

We have performed cyclic sweeps of the compression strain $\varepsilon$ and determined the corresponding pathways and switching fields of sample A over a range of tilt angles $\alpha$ (Fig.~3c-d). We observe three distinct pathways, labeled i, ii and iii, in
the range $-0.57 \degree \leq \alpha \leq
2.86 \degree$ (Fig.~3c). For tilt angle outside this range (which we label regime 'iv') one or more of the defect beams no longer exhibit sharp snapping transitions, but instead smoothly deform. Hence, they do not act as hysterons. We can detect this loss of sharp transitions and hysteresis by the absence of sharp peaks in the image differences, $\Delta$, and attribute it to the increasing presence of shear deformations when $|\alpha |$ is large  ---  for more discussion on shear, see below.
We note that all three t-graphs are of the Preisach type, meaning that there are no avalanches, and that the sequence of hysterons switches is state independent \cite{martinPRE,mungan2019structure,terzi2020state}. In particular we note that for all pathways, the snapping sequence is the same:
first hysteron 3 flips $0\!\rightarrow\!1$, then hysteron 2, and finally hysteron 1, yielding
a sequence of states $S$: $\{000\}\!\rightarrow\!\{001\}\! \rightarrow\!\{011\}\! \rightarrow\!\{111\}$. However, decompressing from state $\{111\}$, we observe different unsnapping sequences in regime i, ii and iii (Fig.~3c).
We note that while in regime i, a single sweep allows to determine all transitions,
in regime ii and iii additional driving cycles are needed to establish all transitions.

In each pathway, we have determined the values of the switching fields, and plot these as function of $\alpha$ (Fig.~3d). We find that the critical switching fields of a given hysteron vary smoothly with $\alpha$, and that the ordering of the upper switching fields $\varepsilon_i^+$ remains the same in regime i-iii, as expected.
Hence, the ordering of the switching fields is consistent with the existence of the three pathways shown in Fig.~3c.
 Moreover, the lower switching fields
$\varepsilon_i^-$ cross at the boundaries between regime's i-iii. Here,  two hysterons change state at the same strain, which could look like an avalanche. However, we notice that such ``avalanches'' are not robust to small changes in the tilt angle, and thus can be seen as degeneracies \cite{hadrien}.

We note that the broad trends in the variation of the switching fields can be understood geometrically: in lowest order one expects an increase in $\alpha $ to increase the switching fields of hysteron 1, and lower those of hysteron 3. Moreover, the actual trends are more complex,
due to the increasing role of shear (see right panel Fig.~3b) that becomes coupled to compression for $\alpha\ne 0$, and which we have observed to have a strong impact on the hysterons.
Moreover, we also note that the sample is prone to global buckling, which breaks left-right symmetry, and that tilting couples to this instability and leads to shearing in the center region, which affects the behavior of the beams in each hysteron.

We measured the switching field for a given hysteron starting from two distinct states (e.g., $\varepsilon_2^+(001)$ and $\varepsilon_2^+(101)$) in regimes ii and iii, and these give insight into the presence of hysteron interactions.
In the absence of hysteron interactions, the switching fields for a given hysteron should be state independent; the small but systematic deviation between $\varepsilon_2^+(001)$ and $\varepsilon_2^+(101)$ indicates the presence of hysteron interactions, which however do not lead to t-graphs that are more complex than Preisach graphs (Fig.~3d) \cite{martinPRE,mungan2019structure,terzi2020state}.

We conclude that tilting of one of the boundaries allows to elicit multiple
pathways from a single sample, and that the variation of the individual switching fields
both gives an interpretation to the emergences of these pathways, as well as
an experimental tool to probe  hysteron interactions.

\section{Non-hysteron degrees of freedom}\label{sec:memory}

For a system to be described as a collection of (interacting) hysterons, the switching fields can only depend on the current collective state, but not on other aspects of the driving history \cite{martinPRE,lindeman2021multiple,keim2021multiperiodic}.
However, many materials when driven
repeatedly can evolve in different manners, for example as they
suffer from fatigue and plastic aging or exhibit visco-elastic effects \cite{lahini2017nonmonotonic,keim2019memory,yoavpre,rate_dep_meta1,rate_dep_meta2,rate_dep_meta3}. The presence of such additional degrees of freedom can for example be seen in the pathways of crumpled sheets that are cyclically driven \cite{yoavpre}.
As we show below, our samples also feature such additional degrees of freedom, with the relative simplicity of our system allowing us to control, reset and understand these effects.

\begin{figure*}[ht]
	\includegraphics[width=0.8\linewidth]{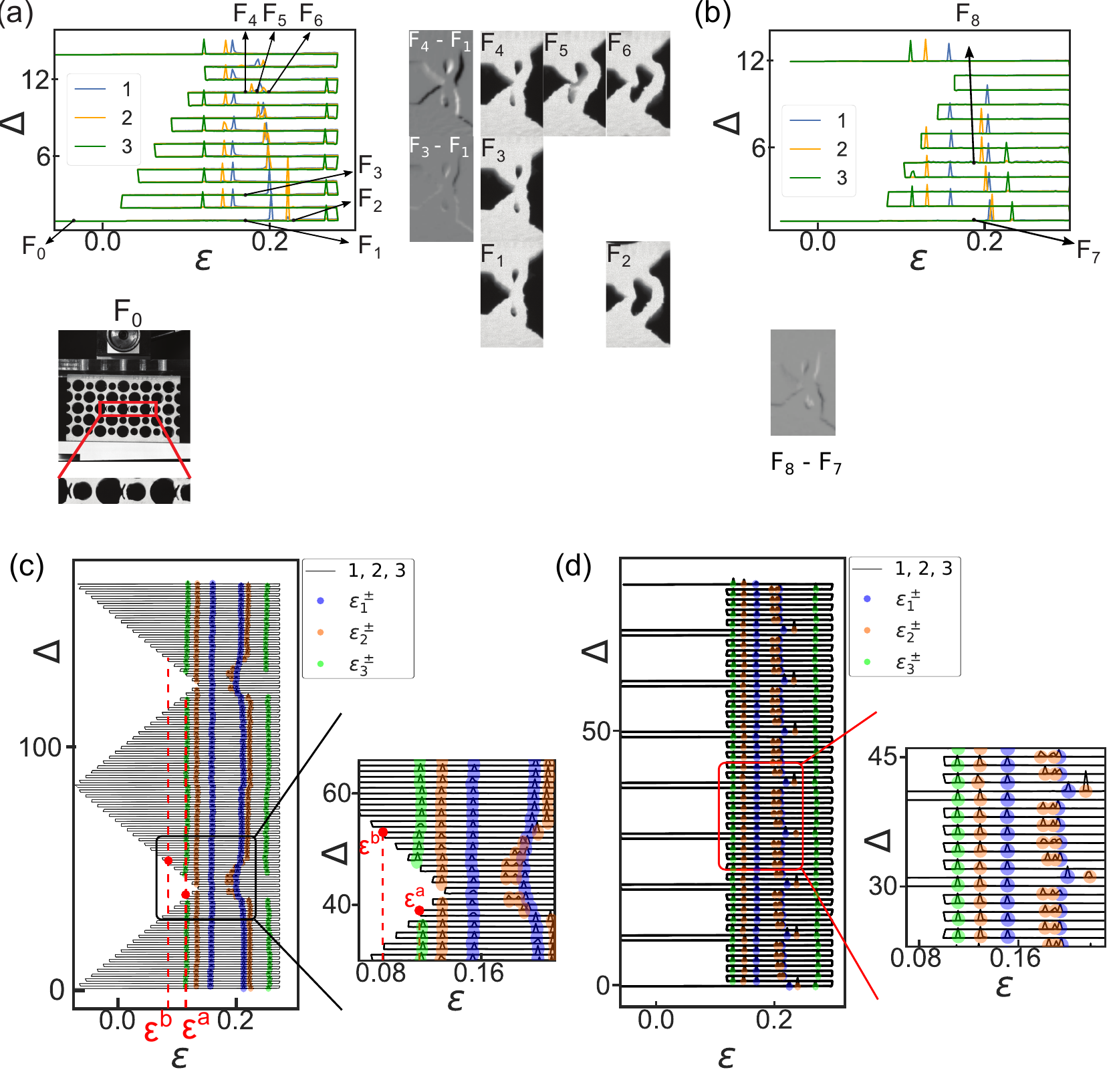}
	\caption{Snapping in sample B and additional degree of freedom.
		(a) The traces of the image difference $\Delta$ for multiple driving cycles where we increase the minimal compression $\varepsilon^m$ as $0.019, 0.039,  0.059, 0.079, 0.099, 0.119$
		(traces offset for clarity). Here $\alpha$ = 0.57$\degree$.
		The location of the spikes on the upsweeps indicates the values of $\varepsilon_i^+$, and we observe the variation of $\varepsilon_1^+$ and $\varepsilon_2^+$ in later cycles. The labels $F1-F6$ indicate the strains and sweeps where we took snapshots (insets).
		The figures $F3-F1$ and $F4-F1$ are difference images, which confirm that while F1 and F3 are nearly identical, F1 and F4 are distinct.
		(b) Image differences for $\alpha$ = -0.29$\degree$ shows
		a weaker evolution of the switching fields  $\varepsilon_1^+$ and $\varepsilon_2^+$.
		(c-d) Sample B, now covered in powder to reduce friction and sticking, and $\alpha$ = 0.57$\degree$.
		Slow sweeps of $\varepsilon^m$ evidence the presence of two critical compressions,
		$\varepsilon_a$ and $\varepsilon_b$ (c), while
		repeated driving at low and high values of $\varepsilon^m$ evidences the absence of
		plasticity, and the presence of an additional degree of freedom with complex dynamics (d). For details, see text.
	}
	\label{fig-sampleB}
\end{figure*}

We probe the presence of additional degrees of freedom by cyclic driving protocols, where we vary the minimum strain, $\varepsilon^{m}$ over time, while probing the switching of each hysteron by monitoring the image differences. To visualize potentially slow evolution, we plot the traces of $\Delta$ as function of strain, offsetting each up and down sweep by one, as in Fig.~1c.
We introduce a new sample B, with $\{u_1, u_2, u_3, h_1, h_2, h_3\} = \{0.2, 0, -0.1, 0.333, 0.300, 0.267\}$, i.e., with the same values of $u_i$ as sample A, but different values of $h_i$.
The larger values of $h$ means pushers come into contact at lower strains than in sample A, which leads to the emergence of an additional frictional degree of freedom.
Indeed, sample B is prone to deviations from purely hysteron-driven behavior, and we study it now in detail.

A first example of such a sweep is shown in Fig.~4a. Here we increase $\varepsilon^{m}$ while keeping the maximum strain $\varepsilon^{M}$ constant. We can clearly observe evolution of the upper switching fields $\varepsilon_1^+$ and $\varepsilon_2^+$, which are lowered in later cycles,
and which even interchange their ordering. We stress that this happens in cycles where
$\varepsilon^{m}$ is low enough so that we return to state $\{000\}$; hence the evolution of $\varepsilon_1^+$ and $\varepsilon_2^+$ goes beyond simple hysteron interactions, and evidences additional degrees of freedom.

A closer inspection of the traces shows that the peak in $\Delta$ corresponding to the switching of hysteron 2 splits in two peaks --- hence instead of a single snapping event, the beam undergoes two discontinuous deformations at two nearby yet distinct values of $\varepsilon$. To see this in more detail, we have monitored the configuration of beam 2 at given fixed $\varepsilon= 0.181$ at various sweeps
(Fig.~4a inset).
These show a clear snapping in the first sweep (frames F1, F2), and
that the state of the beam in the second and first up sweep (F1, F3) are very close, as further evidenced by the absence of a clear signal in the image difference (F1-F3). However, the state of the beam on the upsweep at $\varepsilon=0.039$ slowly evolves when the sweeps are repeated, and indeed F4 is visibly different from F1, as clearly evidenced in their image difference.
In particular, we notice the emergence of shear: while the top and bottom pusher are vertically aligned in the first sweep (F1), after a few sweeps we observe misalignment and shear (F4).
We find that the snapping of such a sheared beam breaks up in two events
(F5 and F6), as also evidenced by the splitting of the relevant peak of $\Delta$ in two separate peaks. We believe that this shear is the main driving force between the shift of $\varepsilon_1^+$ and $\varepsilon_2^+$ that occurs long before such splitting occurs. Consistent with this,
experiments performed at a lesser tilt angle show a similar but weaker evolution of the
switching fields (Fig.~4b).

We now investigate whether shearing is
slaved to the amount of compression, and potentially the hysteron state, or whether it represents (a set of) independent degrees of freedom. Moreover, we will disentangle the role of
visco-plastic effects, friction and stickyness.
To control
the latter, we cover the samples in (baby) powder, which virtually eliminates sticking and lowers the friction, and subject the sample to a large number of sweeps with slowly varying $\varepsilon^m$ (Fig.~4c).
The behavior of the sample depends sensitively on the value of $\varepsilon^m$, and we define two
critical values, $\varepsilon^a=0.11 \pm 0.01 \approx \varepsilon_3^-$ and a smaller value $\varepsilon^b=0.08 \pm 0.01$, where  the errorbar is caused by the increment of $\varepsilon^m$ (Fig.~4c).
The first striking observation is that the evolution of $\varepsilon_1^+$ and $\varepsilon_2^+$ is virtually absent for sweeps where $\varepsilon^m<\varepsilon^a$. This suggests that
visco-plastic effects are not the sole or main driving force, and that lowering the stickyness and friction is important.
Then, when $\varepsilon^m$ is increased beyond  $\varepsilon^a$, we observe a rapid change in the
switching fields $\varepsilon_1^+$ and $\varepsilon_2^+$, which eventually cross, after which the peak for hysteron 2 splits into two peaks, as shown before in Fig.~4a. One could easily interpret
the shifting of the switching fields as hysteron interactions, as when
$\varepsilon^m > \varepsilon^a \approx \varepsilon_3^-$, the system does not relax to state $\{000\}$ but instead is in state $\{001\}$ when hysteron 1 and 2 flip from zero to one.
However, the situation is more complex. First, the switching fields $\varepsilon_1^+$ and $\varepsilon_2^+$ evolve with the number of sweeps, without further changes in the hysteron states. Most strikingly, when we lower $\varepsilon^m$ again, the evolution of $\varepsilon_1^+$ and $\varepsilon_2^+$ from their 'baselevel' only stops when $\varepsilon^m = \varepsilon^b < \varepsilon_3^-$, and indeed continues for a few sweeps where the system periodically returns to its $\{000\}$ state at $\varepsilon^m$. Hence, the evolution of the switching fields $\varepsilon_1^+$ and $\varepsilon_2^+$ evidences the presence of an additional degree of freedom, rather than direct hysteron interactions.

To clarify further that the shift of the switching fields is not visco-plastic and not a direct function of the hysteron states, we perform additional experiments where we, in succession, perform one sweep where $\varepsilon^m=0$, so that the system can relax, and four sweeps where
$\varepsilon^m=0.099<\varepsilon^-_3$, so that the system is driven nonlinearly but always resets to state $\{000\}$ at minimum driving. We clearly observe a different behavior of the switching fields: in the latter cycles, the peak of hysteron 2 has split and lies below that of hysteron 1, whereas in the former, the single peak of hysteron 2 lies above that of hysteron 1. Repeating these cycles evidences very little additional evolution; the behavior of the hysterons depends on $\varepsilon^m$, but not on the deeper history of the sample.

Based on the data in Fig.~4, we interpret the existence of non-hysteron degrees of freedom as follows. First, without powder, friction forces and adhesive forces introduce memory dependent contact forces between beams and pushers. For moderate $\varepsilon^m$ these contacts, which break left-right symmetry, drive the persistent emergence of shear in the sample, as seen in the snapshots in Fig.~4a, which modify the switching fields and snapping behavior.
When $\varepsilon^m$ is small enough, all such contacts are broken, the shear is eliminated and the sample relaxes. For powdered samples, stickiness and friction are  reduced, and the attractive and frictional forces between beam and pusher are much reduced, leading to a larger range of $\varepsilon^m$ where the switching fields are independent of $\varepsilon^m$. However, for sufficiently large
$\varepsilon^m$, opposing pushers come into contact and stay in contact over a substantial part of the sweep, with their contacts acting as a frictional memory that directly couples to shear.
Hence, in this case the switching fields $\varepsilon_1^+$ and $\varepsilon_2^+$ depend on $\varepsilon^m$ (Fig.~4c-d). We note that different designs might be explored to minimize such
'memory within memory' effects. However, in many physical systems one would expect additional
degrees of freedom to play a role, and so we consider our specific example of the coupling between a friction/shear degree of freedom and the hysterons to provide a testing ground for investigating such effects.

Finally, we consider how to describe this additional memory effect.
	We distinguish between two aspects. First, depending on the driving history, and in particular the value of $\varepsilon^m$ with respect to $\varepsilon^a$ and $\varepsilon^b$, we find that either the switching fields $\varepsilon_1^+$ and $\varepsilon_2^+$ are constant, or start to slowly evolve. Hence our data evidences the presence of a 'metabit'; when it is in state '0', 	 $\varepsilon_1^+$ and $\varepsilon_2^+$ are constants,
	when it is in state '1',  	$\varepsilon_1^+$ and $\varepsilon_2^+$ continuously evolve with $\varepsilon^m$, in a manner
	which is beyond a hysteron description as it requires a continuous degree of freedom.

We now focus on finding a minimal model
	for the switching on and off of this metabit. We summarize the key features of our data in  Fig.~5a, which summarizes our observations of the on   ($M=1$) or off  ($M=0$) state of the
	metabit for a sequence of five driving cycles (1)-(5).
	First, we note that our data can only detect a sensitivity to the value of $M$
	during upward sweeps of $\varepsilon$ in the range of $\varepsilon_1^+$ and $\varepsilon_2^+$, which we take as
$\varepsilon_1^M \le \varepsilon \le \varepsilon_2^M$ (indicated with a thick bar).
	Second, we note that a single bit-like degree of freedom
	is not sufficient to describe the evolution of $M$: both $\varepsilon^a$ and $\varepsilon^b$ must play a role, but we see that for $\varepsilon$ larger than both $\varepsilon^a$ and $\varepsilon^b$, $M$ can be both zero or one. In other words, since a single sweep is not sufficient to reach $M=1$ --- rather, we need to sweep up, sweep down, and then sweep up again - capturing the evolution of the metabit requires more than a single binary degree of freedom.

We now show that our data is consistent with the scenario sketched in Fig.~5b, where we use four additional states ${ \{\tilde S}\}=\{A,B,C,D\}$, where $M=1$ in state $C$ and $D$. We stress that these states are independent from the state of the hysterons 1, 2 and 3.
The initial state $A$ has $M=0$, and this is where the system returns for small $\varepsilon$.
The transition from state $A$ to $B$ at $\varepsilon^X$ needs to happen on the up-sweep. We assume that state $B$ has  $M=0$, which requires $\varepsilon^X > \varepsilon^M_2$ (otherwise $M$ could be one  in the relevant region for all upsweeps).
	We then assume that there is a transition from state $B$ to state $C$ with $M=1$ on the down sweep at $\varepsilon^Y$, which implies that  $\varepsilon^a<\varepsilon^Y<\varepsilon^X$.
We stress that while in this scenario $M=1$	on (part of) the down sweeps, this does not modify $\varepsilon_1^+$ and $\varepsilon_2^+$, which only play a role on the upsweeps.
Now two things can happen: if $\varepsilon$ falls below $\varepsilon^a$, the system resets and returns to state A (sweep 2); but if $\varepsilon$ remains above $\varepsilon^a$, state $C$ must switch to a state $D$ with $M=1$, with state $D$ only resetting back to state $A$ for $\varepsilon<\varepsilon^b$. Hence, $\varepsilon^Z>\varepsilon^Y$. Following the transitions,
we find that $M=1$ on sweeps 3 and 4, and only resets to $M=0$ on sweep 5, as required.
 Hence, the presence of two distinct state $C$ and $D$ which have $M=1$ encode the observed scenario (Fig.~5).

To clarify the connection of states $A-D$ with the presence of the metabit $M$, we can also denote them as 0, $\bar{0}$, 1,  and $\bar{1}$ respectively; both states 1 and $\bar{1}$ have $M=1$, but they differ in the value of $\varepsilon$ where they relax to state $A$. We then can interpret the presence of the bar as an additional binary switch, which shows that a combination of two binary degrees of freedom is sufficient to describe the evolution of the metabit $M$ with $\varepsilon$.	We believe that this is the simplest possible scenario consistent with our data, showing the complexity of these memory effects.

\begin{figure}[t!]
	\includegraphics[width=.9\linewidth]{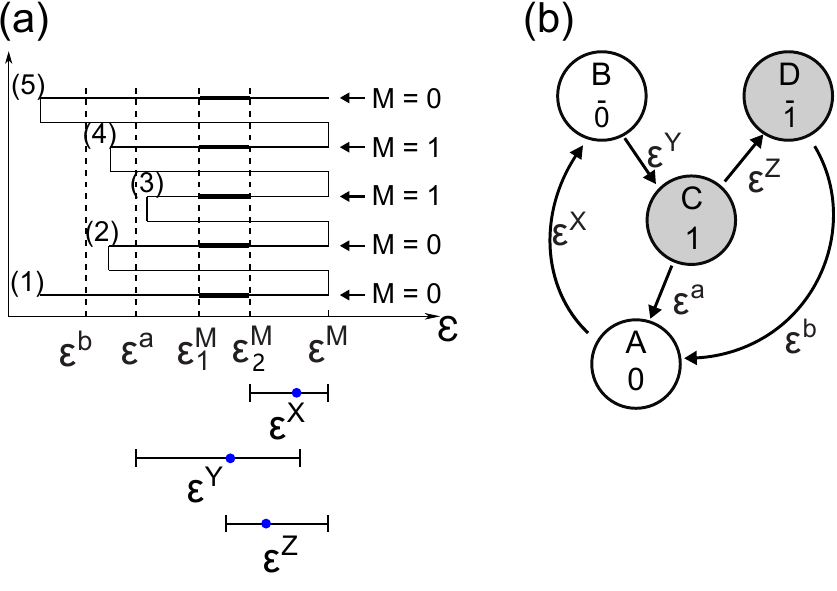}
	\caption{(a) Scenario for the switching on and off of the modifying state $M$. The value of $M$ in the range $\varepsilon^M_1 <\varepsilon< \varepsilon^M_2$ (bold) is indicated for upsweep (1)-(5). (b)
			Tentative state diagram for additional degrees of freedom beyond hysteron 1-3. The system starts out in state A, and only after a sequence of up, down and up transitions, the system can reach the modifying state D, which  leads to changes in the switching fields of hysterons 1 and 2.}
	\label{fig5}
\end{figure}

\section{conclusion}\label{sec:conclusion}

In this paper we have introduced a strategy to embed mechanical hysterons into a metamaterial,
and studied the ensuing pathways under cyclic compression.
We showed how the hysteron properties and pathways can be tuned by the geometric design of
the hysterons, and how the pathways of a given sample can be modified by tilting one of the boundaries. We investigated beyond-hysteron degrees of freedom that modify the switching fields of the hysterons. Our work is a step towards rational design of hysterons and pathways into
metamaterials \cite{martinPRE,Coulais,hadrien}, and moreover highlights the importance of additional degrees of freedom \cite{yoavpre}. Further work may extend these ideas into metamaterials where such additional degrees of freedom can be
controlled, suppressed or leveraged. Moreover, we suggest that alternative designs of mechanical hysterons may allow to tune their switching fields over a wider range. Finally, we are working
on methods to tune the interactions between hysterons, which can extend the range of realizable pathways dramatically \cite{martinPRE}. We hope our
work will inspire further studies on designer matter with targeted pathways.

\section{Supplementary Material}
See supplementary material for experiment procedure and numerical process.

\section{ACKNOWLEDGMENTS}

We thank an anonymous referee for suggestions. M.v.H. acknowledges funding from European Research Council Grant ERC-101019474. J.D. acknowledges kind support from the China Scholarship Council (CSC, No. 201804910512).

\bibliography{JiangnanPaperRef}

\end{document}


\centerline{\textbf{\Large Supplementary Material}}

\section{Experimental Procedure}\label{sec:exp}
Our samples have the well-known biholar geometry, with the essential
geometric parameters described in the main text. We create these samples by 3D printing molds,
and filling these with a two-component elastomer (Zhermack Elite Double 8, $E \approx$ 220 kPa, $\nu \approx$ 0.5).
We use thick samples ($T=18$ mm) to prevent out-of-plane buckling. To ensure homogenous loading,
we extend the top and bottom parts by $P/2$; the effective height of the sample,
$H$, defined as $m \times P$, where $P$ is the pitch of the array and $m$ the number of holes in the vertical direction.

We place our sample in
dual column Instron 5965 uniaxial compression device (resolution better than 4 $\mu$m) and measure the
compressive forces (using a 100 N load cell yielding a resolution better than 0.5 mN) during
cycling loading between compressive strains
$\varepsilon^{m}$ and $\varepsilon^{M}$. We use a (de)compression rate of 0.2 mm/sec to minimize
inertial, viscous and plastic effects.
The rubbers are known to exhibit both visco-elastic and minor plastic creep. These lead to 		minor hysteresis effects, such as those  		seen  in the stress strain curve for experiments  without snapping behavior, i.e. Fig.~1b when $\varepsilon$ < 0.1.

The sample is clamped between the ground plate and top plate.
The bottom plate can be tilted by $|\alpha|$ < $\pm$ 0.5$\degree$.
To allow the sample to relax, we do not fix the top plate to the sample.
To calibrate the zero point of the strain, we use a linear fit to the force vs compression curves
for $\sigma$  in the range between 0.1 kPa and 0.35 kPa (the strains here are a few percent).

We monitor the compressive force and in parallel use video imaging to capture the sample deformations as function of strain, using a Basler asA2040-25gm/gc CMOS camera (
2048 $\times$ 2048 pixels, framerate 2 fps).
To detect snapping events, we use difference imaging where
$\delta$ is defined as
\begin{equation}\label{eq:M2-appr-DiffPhoto}
	\delta_k =  \sum_{i,j} (\mathbf{I}_{k+1}(i,j) - \mathbf{I}_{k}(i,j) )^2~,
\end{equation}
where $\mathbf{I}_{k}(i,j)$ denotes the pixel array of frame $k$.
To normalize these differences, we use $\Delta$:
\begin{equation}\label{eq:M2-appr-DiffPhoto-Delta}
	{\Delta_k}=  \frac{\delta_k}{<{\delta_k>} },
\end{equation}
where $<\bm{\delta}>$ denotes the mean.

\section{Numerical process}\label{sec:supplement-abaqus}
\begin{figure}[h t]
	\includegraphics[width=0.9\linewidth]{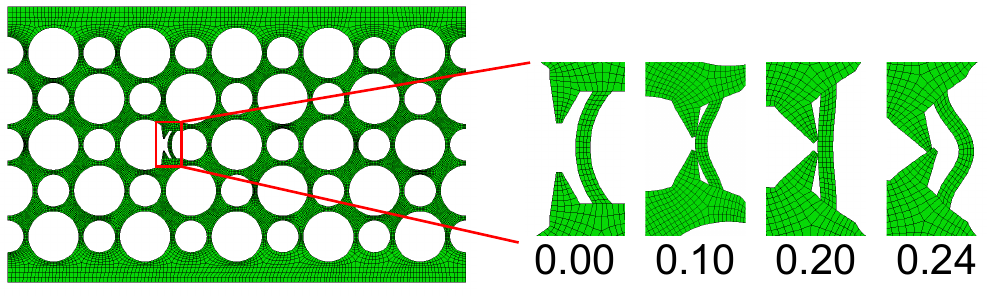}
	\caption{Example of a numerical sample and the evolution of the defect beam geometry under compression, where the sample has same geometry as the sample shown in Fig.~1 (a-b).}	
	\label{fig6}
\end{figure}

We used Abaqus FEM simulations to investigate the snapping behavior of individual hysterons, as shown in Fig.~\ref{fig6}.
We use a 2D structure, and model the rubber metamaterial with neo-Hookean hyper elastic model, using 4-node bilinear plane strain quadrilaterals, reduced integration and hourglass control.
We have performed a systematic mesh refinement study for the in-plane grid, leading to an optimal mesh size of $t/3$.
The parameters of our nearly incompressible isotropic neo-Hookean material, $C_{10}$ and $D_1$, are given by the shear modulus and bulk modulus, $\mu$ and $K$, which are described by Poisson's ratio of the rubber ($\nu$ = 0.495) and the Young's modulus ($E$ = 220 KPa).
\begin{equation}\label{eq:M2-appr-C10-D1}
	\begin{array}{c}
		C_{10} =\frac{1}{2} \mu  \\
		D_1 = \frac{2}{K} \\
		\mu = \frac{E}{2 (1+\nu)}\\
		K = \frac{E}{3 (1- 2\nu)}\\
	\end{array}
\end{equation}
To dissipate the kinetic energy caused by snapping, we apply damping to the model, and the critical damping factor:
\begin{equation}\label{eq:M2-appr-damping}
	C_{10} =\frac{\alpha_R}{2\omega_i} + \frac{\beta_R \omega_i}{2}  \\
\end{equation}
where  $\omega_i$ is the natural frequency; $\alpha_R$ = 0.1; $\beta_R$ = 0.
We have verified that the choices $\alpha_R=0.1$ and $\beta_R=0$ damp out spurious oscillations, do not slow down the simulations significantly, and that the results do not sensitively depend on this choice.

To observe the action of the defect beam, we apply compression on the top surface of the sample and fix the bottom surface, track the center part of the defect beam $u_b$, thus obtaining
time traces $u_{b}(\varepsilon)$. We detect the  snapping and unsnapping strain by finding peaks
in $u_{b}(\varepsilon)$, which we do by requiring that
its derivative exceeds ten times its mean value.

